\documentstyle[multicol,aps,prl,epsf]{revtex}
\begin{document}
\draft
\title{Bound states and extended states around a single vortex in the $d$-wave
superconductors}
\author{Masaru Kato}
\address{Department  of Mathematical Sciences, Osaka Prefecture University,
Sakai, Osaka 599-8531, Japan}
\author{Kazumi Maki}
\address{Department of Physics and Astronomy, University of Southern California,
Los Angeles, CA 90089-0484, USA}
\date{\today}
\maketitle
\begin{abstract}
Making use of the Bogoliubov-de Gennes equation for the $d$-wave superconductors,
we investigate the quasi-particle spectrum around a single vortex.
Taking $p_F\xi=10$,
we found that there are bound states which are localized around the vortex core,
and extended states which are rather uniform, 
for $\left|E\right|<\Delta$ where $E$ is the quasi-particle energy and $\Delta$ is
the asymptotic value of the order parameter for away from the vortex.
\end{abstract}

\begin{multicols}{2}
\narrowtext
Much attention has been paid to the quasi-particle spectrum around a single
vortex in $d$-wave superconductors,
since the discovery of the high-$T_c$ cuprate
superconductors \cite{soininen,maki-schopohl,wang}.
As is well known, $d$-wave superconductivity in both the
hole-doped and the electron doped high-$T_c$ cuprates
has been established \cite{harlingen,tsuei,tsuei2,kokales,prozorov}.

Recent scanning tunneling spectroscopy (STS) sees
different core structures of the vortex in YBCO and Bi2212
\cite{maggio,renner,pan}.
Contrary to the theoretical expectation \cite{soininen,maki-schopohl,wang},
a) they observed a bound state with energy $E$, a fraction of $\Delta$,
and b) they have not seen any clear fourfold symmetry.

Motivated by these experiments,
Morita \textit{et al.} \cite{morita} have abondoned
the quasi-classical approximation used in earlier analysis \cite{maki-schopohl}
and proposed to study the bound states in terms of Bogoliubov-de Gennes
equation \cite{caroli}.
Indeed, this approach appeared to give the correct description of the observation.
However, later Franz and Te\v{s}anovi\'{c} claimed that
there should be no bound states \cite{franz}.
Further this claim was confirmed by Yasui and Kita\cite{yasui}
and by Takigawa \textit{et al.} \cite{takigawa} later.
In a vortex in $s$-wave superconductors, Caroli \textit{et al.} \cite{caroli}
have shown there are a series of bound states.
Further a detailed structure of the bound state wave function is explored
later by Gygi and Schl\"{u}ter \cite{gygi}.
As we have shown later, there are many bound states
in $d$-wave superconductors
as in $s$-wave superconductors \cite{mkkm1}.
We don't know for sure the origin of this disagreement.
But the possible source of their errors
in previous studies \cite{franz,yasui,takigawa} is easy to locate.
They neglected in their analysis the conservation of the angular momentum
around the vortex.
Of course the strict conservation of the angular momentum is broken due to
the fourfold symmetry of $\Delta\left(\bbox{k}\right)$.
On the other hand the angular momentum is still conserved
by modulo 4, and this is adequate to gurantee the presence
of bound states.
Therefore the structure of the bound states in the vicinity of the vortex core
of $d$-wave superconductor appear to be very similar to
the one in $s$-wave superconductor.
Then the only clear difference is the presence of extended states first discovered
in \cite{morita}.
Also these extended states give rise to the Volovik effect\cite{volovik},
which has been observed as the $\sqrt{H}$-term in the
specific heat\cite{moler} and more recently as the $H$ linear term
in the thermal conductivity\cite{kubert,taill}.

Earlier we have considered extreme quantum limit,
\textit{i.e.} $p_F\xi\simeq 1$ \cite{mkkm1}, but recent experiment shows
$p_F\xi\simeq 20,\ 30$ for Bi2212 and YBCO, respectively\cite{taill}.
Therefore, in this letter, we take $p_F\xi=10$.
We consider quasi-two dimensional $d_{x^2-y^2}$-wave superconductors
in a disk with radius $R$.
We put a single vortex at the center of the disk.
When $R$ is large enough, the magnetic field is weak and the vector potential
is small, so we ignore the vector potential.
The Bogoliubov-de Gennes equations are given as,
\begin{eqnarray}
\left[-\frac{1}{2m_e}\nabla^2-\mu\right]u_n\left(\bbox{r}\right)
\nonumber\\
-\frac{1}{p_F^2}\left[\partial_x\Delta\left(\bbox{r}\right)\partial_x
-\partial_y\Delta\left(\bbox{r}\right)\partial_y\right]v_n\left(\bbox{r}\right)
=E_nu_n\left(\bbox{r}\right),\\
-\left[-\frac{1}{2m_e}\nabla^2-\mu\right]v_n\left(\bbox{r}\right)
\nonumber \\
-\frac{1}{p_F^2}\left[\partial_x\Delta\left(\bbox{r}\right)\partial_x
-\partial_y\Delta\left(\bbox{r}\right)\partial_y\right]u_n\left(\bbox{r}\right)
=E_nv_n\left(\bbox{r}\right).
\end{eqnarray}
The order parameter is given as,
\begin{eqnarray}
\Delta\left(\bbox{r}\right)=
\sum_n\frac{1}{p_F^2}\left(1-2f\left(E_n\right)\right)
\nonumber\\
\times\left[\left(\partial_xu_n\left(\bbox{r}\right)\right)
\left(\partial_xv_n^* \left(\bbox{r}\right)\right)
-\left(\partial_yv_n\left(\bbox{r}\right)\right)
\left(\partial_yu_n^* \left(\bbox{r}\right)\right)\right].
\end{eqnarray}
In the following we assume that the vector potential takes following form,
\begin{equation}
\Delta\left(\bbox{r}\right)=\left|\Delta\left(r\right)\right|e^{-i\theta}.
\end{equation}
We use the Fourier-Bessel expansion following Gygi-Schl\"{u}ter as,
\begin{eqnarray}
u_n\left(\bbox{r}\right)=\sum_{m=-\infty}^\infty\sum_{j=1}^\infty
u_{nmj}\phi_{mj}\left(r\right)\frac{e^{im\theta}}{\sqrt{2\pi}},\\
v_n\left(\bbox{r}\right)=\sum_{m=-\infty}^\infty\sum_{j=1}^\infty
v_{nmj}\phi_{mj}\left(r\right)\frac{e^{im\theta}}{\sqrt{2\pi}},
\end{eqnarray}
where $\phi_{mj}$ is Fourier-Bessel basis given as,
\begin{equation}
\phi_{mj}\left(r\right)=\frac{\sqrt{2}}{RJ_{m+1}\left(\alpha_{mj}\right)}
J_m\left(\frac{\alpha_{mj}r}{R}\right).
\end{equation}
Here $\alpha_{mj}$ is the $j$-th zero of the Bessel function of the $m$-th order.
Then the Bogoliubov-de Gennes equation becomes as,
\begin{eqnarray}
\left[\frac{1}{2m_e}\left(\frac{\alpha_{mj}}{R}\right)^2-\mu\right]u_{nmj}
&-&\frac{1}{2}\sum_j^\infty\left(\Delta_{mm-1}^{jj'}v_{nm-1j'}\right.
\nonumber\\
\left.+\Delta_{mm+3}^{jj'}v_{nm+3j'}\right)&=&E_nu_{nmj},\label{eq:BdG2-1}\\
-\left[\frac{1}{2m_e}\left(\frac{\alpha_{mj}}{R}\right)^2-\mu\right]v_{nmj}
&-&\frac{1}{2}\sum_j^\infty\left(\Delta_{mm-1}^{jj'}u_{nm-1j'}\right.
\nonumber\\
\left.+\Delta_{mm+3}^{jj'}u_{nm+3j'}\right)&=&E_nv_{nmj}.\label{eq:BdG2-2}
\end{eqnarray}
$\Delta_{mm-1}^{jj'}$'s are matrix elements and given as,
\begin{eqnarray}
\Delta_{mm-1}^{jj'}=\int_0^Rrdr\Delta\left(r\right)
\phi_{mj}^d\left(r\right)\phi_{m-1j'}^i\left(r\right),\\
\Delta_{mm+3}^{jj'}=\int_0^Rrdr\Delta\left(r\right)
\phi_{mj}^i\left(r\right)\phi_{m+3j'}^d\left(r\right),
\end{eqnarray}
where $\phi_{mj}^d\left(r\right)=\frac{\sqrt{2}}{RJ_{m+1}\left(\alpha_{mj}\right)}
J_{m-1}\left(\frac{\alpha_{mj}r}{R}\right)$ and
$\phi_{mj}^i\left(r\right)=\frac{\sqrt{2}}{RJ_{m+1}\left(\alpha_{mj}\right)}
J_{m+1}\left(\frac{\alpha_{mj}r}{R}\right)$.
The order parameter is now given by,
\begin{eqnarray}
\Delta\left(r\right)=-\frac{g}{4\pi}\sum_n\left[1-2f\left(E_n\right)\right]
\sum_{j_1j_2}\sum_m
\nonumber\\
\left[\phi_{mj_1}^i\left(r\right)\phi_{m+3j_2}^d\left(r\right)u_{nmj_1}v_{nm+3j_2}
\right.\nonumber\\
\left. +\phi_{mj_1}^d\left(r\right)\phi_{m-1j_2}^i\left(r\right)
u_{nmj_1}v_{nm-1j_2}\right].\label{eq:OP2}
\end{eqnarray}
In order to fix the particle density, we determine the chemical potential by the condition,
\begin{equation}
N_e=2\sum_{njm}\left[f\left(E_n\right)\left|u_{nmj}\right|^2
+\left(1-f\left(E_n\right)\right)\left|v_{nmj}\right|^2\right],
\end{equation}
where $N_e$ is total particle number.

From Eqs.\ \ref{eq:BdG2-1} and \ref{eq:BdG2-2},
there are coupled sequences of $u_m$ and $v_m$ and there are four sectors of them.
\begin{eqnarray}
0{\rm th}\ &\cdots&, u_{n-4j}, v_{n-1j}, u_{n0j}, v_{n3j}, \cdots\nonumber\\
1{\rm st}\ &\cdots&, v_{n-4j}, u_{n-3j}, v_{n0j}, u_{n1j}, \cdots\nonumber\\
2{\rm nd}\ &\cdots&, v_{n-3j}, u_{n-2j}, v_{n1j}, u_{n2j}, \cdots\nonumber\\
3{\rm rd}\ &\cdots&, v_{n-2j}, u_{n-1j}, v_{n2j}, u_{n3j}, \cdots
\end{eqnarray}
There is a following symmetry between these sectors,
\begin{eqnarray}
\left\{u_{nm}, v_{nm}, E_n\right\}_1\leftrightarrow
\left\{-v_{n-m}, v_{n-m}, -E_n\right\}_0,
\nonumber\\
\left\{u_{nm}, v_{nm}, E_n\right\}_2\leftrightarrow
\left\{-v_{n-m}, v_{n-m}, -E_n\right\}_3.
\end{eqnarray}
Therefore we solve Eqs.\ (\ref{eq:BdG2-1}) and  (\ref{eq:BdG2-2})
for the 1st and the 2nd sectors and using the symmetry
we determine $\Delta\left(r\right)$.

For the numerical calculation, we take $R=15\xi$, $\Delta_0/E_c=0.4$
and $p_F\xi=10$.

\begin{figure}
\narrowtext \vskip 0mm
\epsfxsize=0.9\hsize \centerline{\vbox{ \epsffile{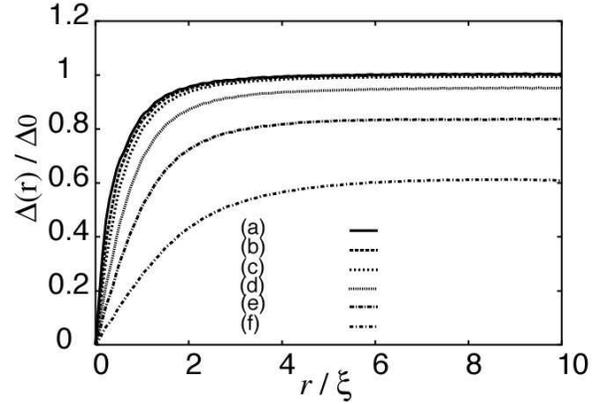} }}
\caption{Temperature dependence of the order parameter.
(a) $T/T_c=0.1$, (b) $T/T_c=0.2$, (c) $T/T_c=0.3$, (d) $T/T_c=0.5$,
(e) $T/T_c=0.7$, and (f) $T/T_c=0.9$.}
\label{fig:OP}
\end{figure}

First, we show the temperature dependence of the order parameter
in Fig.\ \ref{fig:OP}.

There is an oscillation of the order parameter away from the
vortex core.
This comes from the geometrical resonance of the system
and the boundary condition.
Comparing with the $s$-wave superconductors\cite{gygi,mkkm-s},
there is no structure inside of the core.

\begin{figure}
\narrowtext \vskip 0mm
\epsfxsize=.8 \hsize \centerline{\vbox{ \epsffile{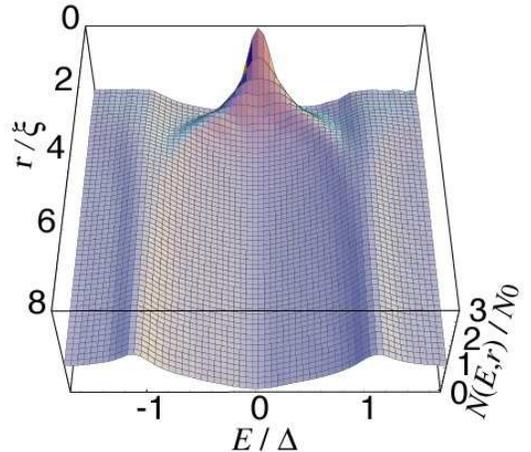} }}
\caption{Local density of states as a function of $E$ and $r$ at $T/T_c=0.1$.
It is normalized by density of states of normal state $N_0$.}
\label{fig:ldos}
\end{figure}
\begin{figure}
\narrowtext \vskip 0mm
\epsfxsize=.8 \hsize \centerline{\vbox{ \epsffile{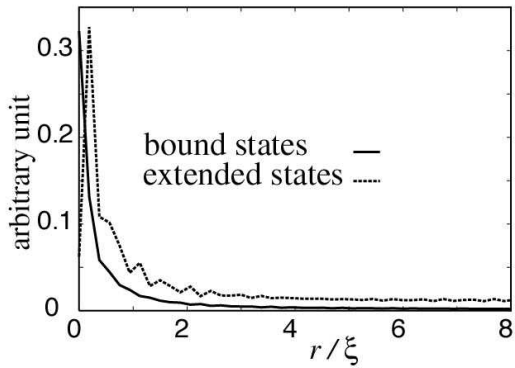} }}
\caption{Weight of bound states and extended states
for $E/\Delta=0.05\pm0.02$.}
\label{fig:eigen}
\end{figure}
\begin{figure}
\narrowtext \vskip 0mm
\epsfxsize=.7\hsize \centerline{\vbox{ \epsffile{fig4.EPSF} }}
\caption{Wave functions of a bound state,
 (a) $\left|u\left(\bbox{r}\right)\right|^2$
and (b) $\left|v\left(\bbox{r}\right)\right|^2$.
Its energy eigenvalue is $E/\Delta=0.0481$.}
\label{fig:bound}
\end{figure}

We show the local density of states for $T/T_c=0.1$ in Fig.\ \ref{fig:ldos}.
There is a small angular dependence,
but we integrate out the angular dependence for clarity.
Because of the nodes of the $d_{x^2-y^2}$ order parameter, 
there is a finite density of states inside of the energy gap.
Also there is a peak around $r=0$ and $E=0$ and
a broad peak close to the boundary.
Our previous calculation shows that for $p_F\xi=1.0$,
the peak around a vortex core moves to gap edge\cite{mkkm1}.
In order to examine the origin of this peak,
we show the relative weight of the bound state versus the extended state
for $E/\Delta =0.05\pm 0.02$
as function of $r/\xi$ in Fig.\ \ref{fig:eigen}.
The relative weight means a sum of $\int_0^{2\pi}
 \left[u\left(\bbox{r}\right)^2+v\left(\bbox{r}\right)^2\right]d\theta$.
As is readily seen the bound states are localized near the vortex core,
while the extended states are rather flat.
So this should give a quasi-uniform density of states at $E=0$ \cite{maggio}.
These are shown is Fig.\ \ref{fig:bound} and Fig.\ \ref{fig:extended},
respectively.
\begin{figure}
\narrowtext \vskip 0mm
\epsfxsize=.7 \hsize \centerline{\vbox{ \epsffile{fig5.EPSF} }}
\caption{Wave functions of an extended state,
(a) $\left|u\left(\bbox{r}\right)\right|^2$
and (b) $\left|v\left(\bbox{r}\right)\right|^2$.
Its energy eigenvalue is $E/\Delta=0.0331$.}
\label{fig:extended}
\end{figure}

Strictly speaking, we find other set of solutions, which cannot be
characterized as the bound state or the extended state.
But we believe that they are due to the finite size of our disk
which we have studied.

\begin{figure}
\narrowtext \vskip 0mm
\epsfxsize=1.0\hsize \centerline{\vbox{ \epsffile{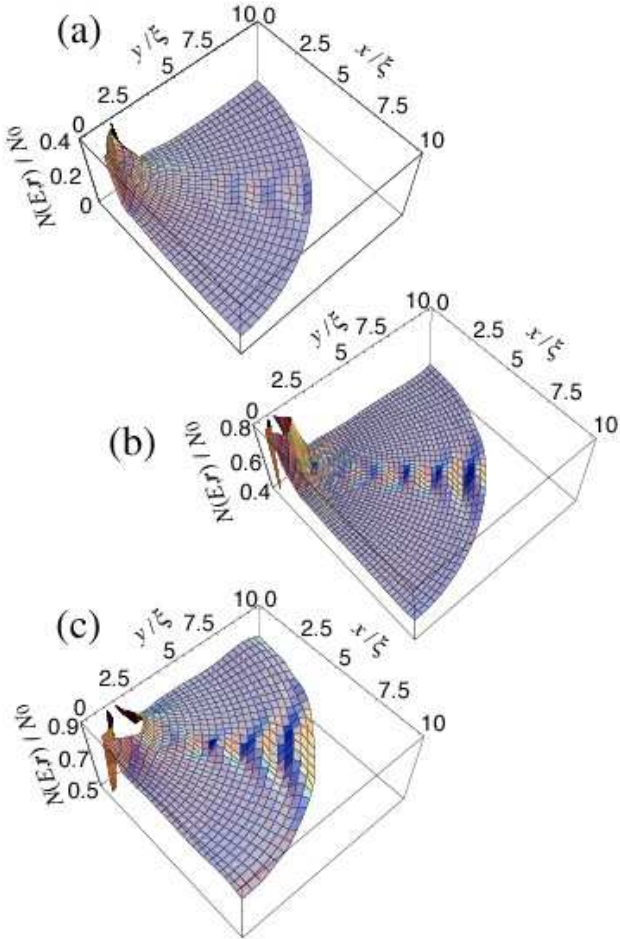} }}
\caption{Angular dependence of the local density of states.
(a) $E=0$, (b) $E/\Delta=0.5$, and (c) $E/\Delta=0.7$.
They are normalized by the density of states of normal state $N_0$.}
\label{fig:ldosangle}
\end{figure}

The angular dependence of the density of states is shown
in Fig.\ \ref{fig:ldosangle}.
There is an oscillation of the LDOS along the diagonal direction ($x=y$).
This oscillation becomes large for large $E$.


In Summary, making use of the Bogoliubov-de Gennes equation
for $d$-wave superconductors \cite{caroli},
we have studied the quasi-particle spectrum around a single vortex
with a weak-coupling model and for $p_F\xi=10$.
First, we find many bound states centered around the vortex core as shown in
Fig.\ \ref{fig:ldos}.
This picture is very similar to the one found
for $s$-wave superconductors \cite{gygi}.
Second, ther are extended states with almost uniform amplitude
near the vortex core (see Fig.\ \ref{fig:extended}).
Third, there appear to be a small group of
the mixture of the bound state and the extended state.
However, we believe this mixing is the finite size effect.
In order to clarify this question, we need a pararell study of the
quasi-particle spectrum in disks with different size.
There remains one mystery.
Why STM in YBCO \cite{maggio} and Bi2212 \cite{pan} picked up
only a single bound state?
Perhaps some of bound states are more readily accessible to STM?

\acknowledgments
One (KM) of us thanks Peter Fulde and Max-Planck Institue f\"{u}r
Physik komplexer Systeme at Dresden for their hospitality where a part of this
work is carried out.
We thank Stephen Haas for useful discussions and suggestions.

\end{multicols}
\end{document}